\documentclass[reprint,
 amsmath,amssymb,
 aps,
 prl,
 superscriptaddress
]{revtex4-1}

\usepackage{graphicx}
\usepackage{dcolumn}
\usepackage{bm}

\begin{document}

\preprint{APS/123-QED}

\title{Strain-dependent solid surface stress and the stiffness of soft contacts}

\author{Katharine E. Jensen}
\altaffiliation{Present address: Williams College Department of Physics, Williamstown, MA, USA; kej2@williams.edu}
\affiliation{
 Department of Materials, ETH Z\"{u}rich, Switzerland
}

\author{Robert W. Style}
\affiliation{
 Department of Materials, ETH Z\"{u}rich, Switzerland
}
\author{Qin Xu}
\affiliation{
 Department of Materials, ETH Z\"{u}rich, Switzerland
}
\author{Eric R. Dufresne}
\altaffiliation{eric.dufresne@mat.ethz.ch}
\affiliation{
 Department of Materials, ETH Z\"{u}rich, Switzerland
}

\date{\today}

\begin{abstract}
Surface stresses have recently emerged as a key player in the mechanics of highly compliant solids.
The classic theories of contact mechanics describe adhesion with a compliant substrate as a competition between surface energies driving deformation to establish contact and bulk elasticity resisting this.
However, it has recently been shown that surface stresses provide an additional restoring force that can compete with and even dominate over elasticity in highly compliant materials, especially when length scales are small compared to the ratio of the surface stress to the elastic modulus, $\Upsilon/E$.
Here, we investigate experimentally the contribution of surface stresses to the force of adhesion.
We find that the elastic and capillary contributions to the adhesive force are of similar magnitude, and that both are required to account for measured adhesive forces between rigid silica spheres and compliant, silicone gels.
Notably, the strain-dependence of the solid surface stress contributes significantly to the stiffness of soft solid contacts.

\end{abstract}

\maketitle

Soft solids can make excellent adhesives because they can conform to establish intimate contact, even on very rough surfaces \cite{CretonPapon2003,Creton2003}. 
Applications of soft or ``pressure-sensitive'' adhesives range from the ubiquitous sticky note to large-scale building construction \cite{Creton2003}, from everyday adhesive bandages to new developments toward improved surgical technique \cite{rose2014nanoparticle}.
The true test of any adhesive material is how it responds to an externally-applied force. 
Does it stick and stay stuck, and how much force can it sustain before unsticking? 
Even though soft adhesives are widely used, answering these seemingly-simple questions remains an area of active research \cite{ShullCrosby1998, Chakrabarti2013, StyleNatComm2013, Salez2013, Xu2014, Cao2014, Hui2015, jensen2015wetting, Liu2015, nalam2015nano, minsky2017composite}. 

When a soft solid conforms into adhesive contact with an uneven surface, it is well understood that bulk elasticity opposes this deformation \cite{JKR1971,JohnsonBook1987,Maugis1995}.
However, a number of recent experiments have demonstrated that for highly compliant solids,  elasticity is not always enough to describe the mechanical response \cite{Long1996, Mora2010, Jerison2011, Jagota2012, StylePRL2013, Nadermann2013, StyleNatComm2013, Paretkar2014, cao2014elastocapillarity, style2015stiffening, andreotti2016soft,cao2016nanoparticles, long2016effects,style2017elastocapillarity}.
Rather, an additional restoring force can arise from the solid surface tension, $\Upsilon$, which opposes the stretching of the surface required to conform into contact.
This solid surface stress can compete with or even dominate over the elastic modulus, $E$, in determining the mechanics of soft materials, at least on length scales that are small compared to an elastocapillary length, $L_c = \Upsilon/E$.

Meanwhile, surface stresses are still ignored in the standard theories of adhesive contact mechanics \cite{JKR1971,JohnsonBook1987,Maugis1995}.  
Recent insights into elastocapillary phenomena suggest that a new approach is needed
to interpret contact measurements on soft materials, from characterizing cancer cells using atomic force microscopy to soft adhesives development \cite{van2003biomechanics,shull2002contact}. 
Theoretical studies have begun to investigate the contributions of surface stresses to adhesive forces \cite{Hui2015,liu2016effect}, but there are not yet experimental data. 

In this paper, we investigate the roles of surface tension and elasticity in adhesion with applied force.
We directly measure the adhesive forces and contact geometry  between compliant solid substrates and small rigid spheres during quasi-static separation.
We find that classic theories of contact mechanics  fail to account for either the forces or the shape of the contact zone.
On the other hand, the measured forces are reasonably described when a simple estimate of the contribution of surface stress is added to the standard elastic predictions.  
We find that the strain-dependence of the sold surface stress plays an essential role in these phenomena.

We study the pull-off of small glass spheres from compliant, silicone gel substrates.
The gels are prepared by mixing liquid (1 Pa$\cdot$s) divinyl-terminated polydimethylsiloxane (PDMS) (Gelest, DMS-V31) with a chemical cross-linker (Gelest, HMS-301) and catalyst (Gelest, SIP6831.2) (as in Ref. \cite{jensen2015wetting, style2015stiffening}).
We degas the mixture in vacuum, and then deposit a layer along the millimeter-wide edge of a standard microscope slide.
After curing at 68$^\circ$C overnight, the resulting solid silicone substrate is about 300 $\mu$m thick, flat parallel to the long edge of the microscope slide, and very slightly curved (radius of curvature $\sim$700 $\mu$m) in the orthogonal direction \cite{jensen2015wetting}.
The cured PDMS substrate has a Young modulus of $E = 5.6$ kPa, and the Poisson ratio of the gel's elastic network is $\nu = 0.48$ \cite{jensen2015wetting,TFMpaper}.
Bulk tensile tests show that the gel is linear elastic to about 10\% true strain and moderately strain stiffening thereafter \cite{xujensen2017direct}.

\begin{figure*}[ht!]
\includegraphics[width = \textwidth]{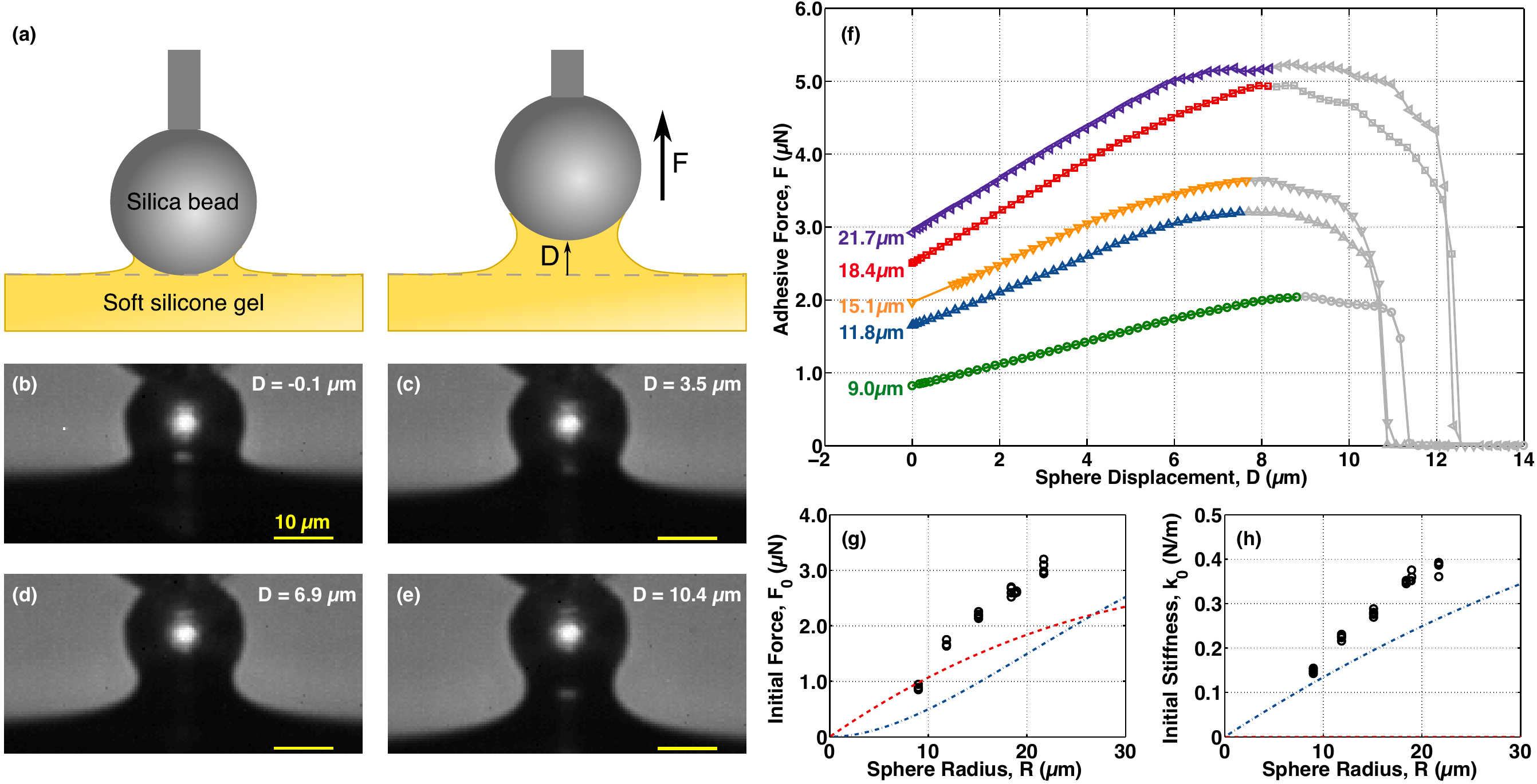}
\caption{\label{force}Structure and force of soft adhesive contacts.  
\textit{(a)} Schematic of the experiment. 
\textit{(b-e)} Raw images of adhesive contact between a 10.0-$\mu$m-radius sphere and an initially-flat, compliant silicone substrate with a Young modulus of $E = $5.6 kPa. (See Supplemental Materials for movies of initial contact and quasi-static pull \cite{supplement}.) 
\textit{(f)} Examples of force-displacement measurements for a range of sphere sizes, as indicated.
Colored points indicated the stable contact regime; gray points show the unstable regime through detachment.
\textit{(g-h)} Initial contact force, $F_0$, and contact stiffness, $k_0$, are plotted versus sphere radius, $R$, respectively. 
Elastic contact mechanics predictions \cite{Maugis1995} are overlaid as blue dot-dashed lines. 
Simple capillary predictions with a fixed value of surface stress, $\Upsilon \approx 0.02$ N/m \cite{jensen2015wetting}, are plotted as red dashed lines.}
\end{figure*}

For rigid, spherical indenters, we use untreated silica spheres ranging in radius from 7.9 to 32.0 $\mu$m (Polysciences, 07668).
We rigidly attach the spheres to the ends of either rigid, tapered glass rods or solid-state capacitive force probes (FemtoTools, FTS 100) using two-part 5-minute epoxy (Elmer's), waiting at least 6 minutes after mixing to ensure that the glue does not flow over the sphere surface.
For the spheres attached to tapered glass rods, we control the position manually with sub-micrometer precision using a 3-axis micromanipulator stage (Narshige MMO-023).
For the spheres attached to the force probes, we control the position using a 3-axis piezo stage with 1-nm accuracy (FemtoTools).
Either indentation system is mounted on a standard inverted microscope, and the contact zone is imaged from the side with a 40x (N.A. 0.60) objective, as described previously \cite{jensen2015wetting,xujensen2017direct}.

We begin each experiment by bringing a sphere into initial adhesive contact with the solid silicone gel substrate at a vertical position $D=0$, where $D$ is defined as the distance between the initial, undeformed surface and the bottom of the sphere, as shown in Figure \ref{force}(a).
We approach slowly until the bottom of the sphere just touches an initially-flat region of substrate that has not been contacted previously.
We  identify contact either as the first position where we register a measurable force, or where we visually observe the compliant substrate suddenly deforming into contact with the sphere.
High-speed imaging indicates that the initial contact deformation is complete in less than a second \cite{supplement}.
The rigid attachment of the sphere prevents it  from spontaneously indenting into the substrate \cite{StyleNatComm2013,jensen2015wetting}, so at the start of the experiment the substrate already exerts an initial tensile force, $F_0$.

We wait about 10 minutes after initial contact in order to ensure the system is in equilibrium before beginning each experiment.
We then quasi-statically withdraw the sphere from the surface ($D>0$) at a slow rate of 0.1 $\mu$m/s. 
A series of example images from a typical experiment on a 10.0-$\mu$m-radius sphere are shown in Figure \ref{force}(b-e).
At initial contact (Figure \ref{force}(b)), we already observe significant local deformation of the substrate. 
As we subsequently pull the sphere away from initial contact, the contact area stays nearly constant, decreasing only slightly as we approach the last stable position (Figure \ref{force}(e)).
After this position, the contact line begins to slide rapidly toward a point at the bottom of the sphere where it finally detaches.
For smaller displacements, the solid adhesive bridge between the bulk of the substrate and the sphere is stable.

Examples of raw force-displacement data from initial contact ($D = 0$) through detachment for several sphere sizes are shown in Figure \ref{force}(f).
Colored points indicate the stable contact regime before detachment begins, identified here as measurements prior to the maximum recorded force.
All force-displacement measurements start with an initial tensile contact force, $F_0$, that increases with sphere size, as shown in Figure \ref{force}(g).
From initial contact, the force then increases linearly with displacement for much of the stable contact regime.
The contact stiffness, $k_0 = dF/dD|_{D=0}$, of this springlike regime also increases with sphere size, as shown in Figure \ref{force}(h).
Varying the displacement rate to be 2x slower or up to 10x faster affected the peak force and the distance to detachment, but had no measurable effect on $F_0$ or $k_0$.
Repeat measurements with the same sphere in different locations are extremely consistent, varying only in the unstable contact regime.

If the measured adhesive forces were due entirely to elastic restoring stresses, we could estimate the total force using classic adhesive contact mechanics as extended by Maugis for large contact radii \cite{Maugis1995}:
\begin{equation}
F_{EL} = \frac{2 E a}{1-\nu^2} \left[-D - \frac{R}{2} + \frac{R^2 - a^2}{4 a} \ln{\frac{R+a}{R-a}}  \right]
\label{maugis}
\end{equation}
We use this relation to calculate elastic theory predictions for both initial contact force, $F_{0,EL}(R)$ and initial contact stiffness, $k_{0,EL}(R)$, overlayed on our data  as blue dot-dashed lines in Figure \ref{force}(g-h).
To generate a smooth curve, we interpolate the measured contact radii (as described in the Supplemental Materials \cite{supplement}).
Since $a$ is only very weakly dependent on $D$, decreasing $\lesssim$10\% over the entire stable contact regime, we approximate it as constant for small $D$ in estimating $k_{0,EL}(R)$.

Elastic theory consistently underestimates both the initial contact forces and the initial contact stiffness.
One could improve the elastic calculation  by accounting for nonlinear elasticity or large deformations using finite-elements \cite{liu2016effect,xu2016effect}. 
Instead, we consider possible contributions to the force from solid surface stresses.

In the absence of a complete elastocapillary adhesion theory, we calculate the capillary force contribution as the integral of the surface stresses,  $\Upsilon$, at the contact line \cite{Hui2015}: 
\begin{equation}
F_{CL} = 2 \pi a \sin(\Theta)  \Upsilon
\label{cap}
\end{equation}
Here $\Theta$ is the angle from the horizontal at which the surface leaves the contact line.
As we have total wetting between the substrate and spheres \cite{jensen2015wetting}, $\sin(\Theta) = a/R$, and Equation \ref{cap} simplifies to \mbox{$F_{CL} = 2 \pi \left( a^2/R \right) \Upsilon$}.
Note that this is the familiar approach to determining the force exerted by a liquid bridge \cite{deGennes2004,butt2009surface}.

We plot the predictions of this capillary theory as red dashed lines in Figure \ref{force}(g-h), again assuming $a \approx a_0$ for small $D$.
For this calculation, we use the measured zero-force surface tension, $\Upsilon_0 \approx 0.02$ N/m \cite{jensen2015wetting}. 
The capillary prediction for the initial force is significant, of the same order of magnitude or larger than the elastic predictions.
This additional restoring force roughly accounts for the entire discrepancy between the measured initial contact forces and the elastic predictions.
However, because Equation \ref{cap} lacks any explicit dependence on sphere displacement, this simple contact line force model does not contribute to the contact stiffness, which remains underestimated.

\begin{figure*}[ht!]
\includegraphics[width = \textwidth]{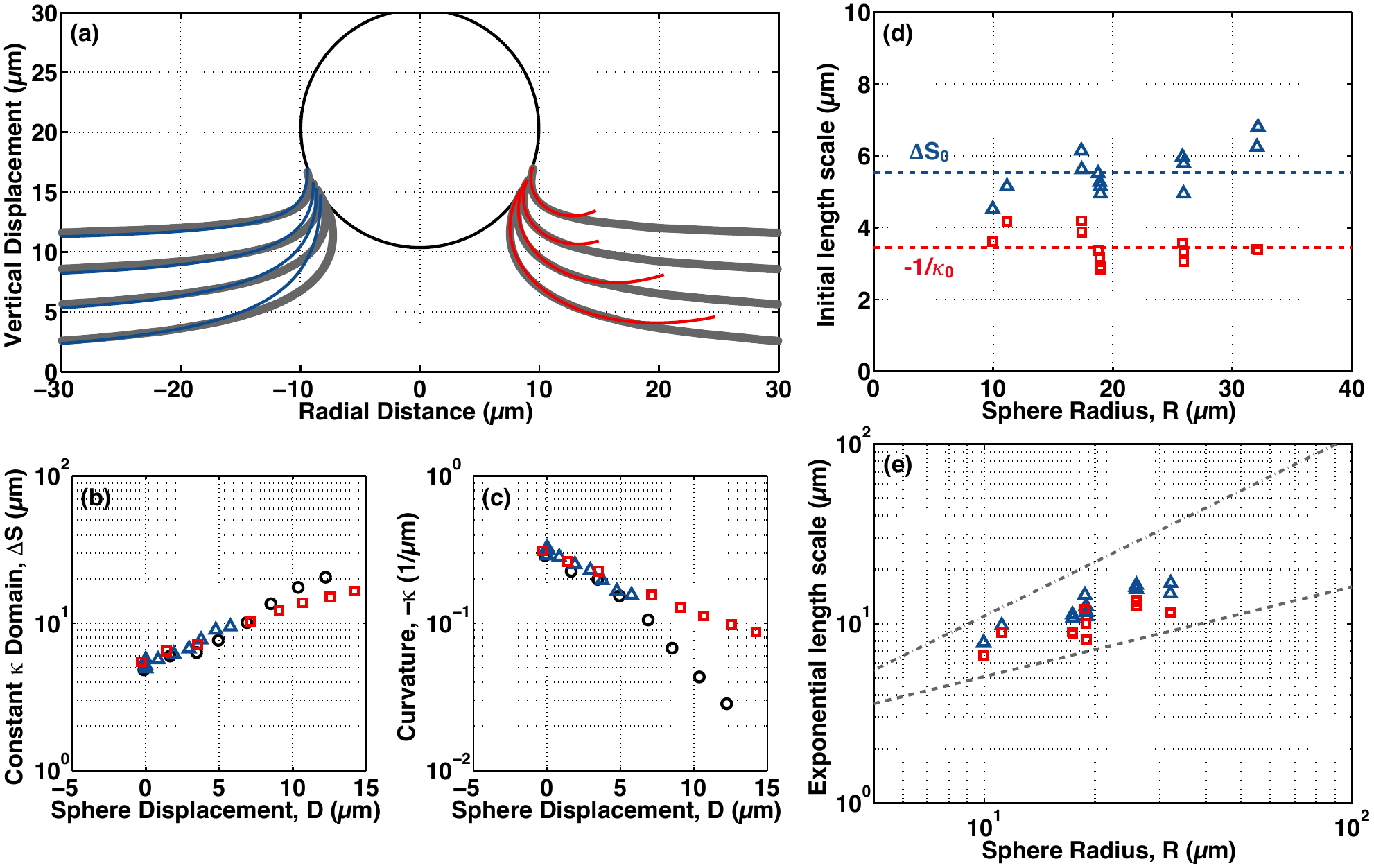}
\caption{\label{geo}Geometry of soft adhesive contact. 
\textit{(a)} Mapped profiles from the same images shown in Figure \ref{force}(b-e) (overlapping gray points). 
Predictions from elastic theory \cite{Maugis1995} are overlaid at left (blue lines).
Constant total curvature fits are overlaid at right (red lines).
\textit{(b)} Size of the domain of constant curvature, $\Delta S$, versus sphere displacement, $D$, for three example experiments with spheres of radius 10.0 $\mu$m (black circles), 18.8 $\mu$m (blue triangles), and 32.0 $\mu$m (red squares). 
\textit{(c)} Total curvature, $-\kappa$, versus $D$ for the same three examples as in (b).
\textit{(d)} Plot of fit initial length scales versus sphere radius, $R$: $\Delta S_0$ (blue triangles, mean $\pm$ std dev $=$ 5.5 $\pm$ 0.6 $\mu$m) and $-1/\kappa_0$ (red squares,  mean $\pm$ std dev $=$ 3.5 $\pm$ 0.4 $\mu$m).
\textit{(e)} Log-log plot of fit exponential length scales, $L_{\Delta S}$ (blue triangles) and $L_{\kappa}$ (red squares) versus $R$.
Lines of slope 1 (dot-dashed) and slope 1/2 (dashed) are shown as guides to the eye.
}
\end{figure*}

To gain more insight into the interplay of elastic and surface stresses during soft adhesion with applied force, we examine the structure of the contact zone.
From the raw image data, we map the dark profile of the sphere and silicone substrate with 100-nm-resolution using edge detection in MATLAB \cite{jensen2015wetting}. 
We also map the undeformed surface before and after the experiment to establish the zero-position of the coordinate system and to check for any permanent deformation or image drift.
Figure \ref{geo}(a) shows the measured substrate surface profiles (overlapping gray points) extracted from the raw image examples in Figure \ref{force}(b-e), shifted so that the sphere remains in a fixed position (black circle).
Since the deformation is axisymmetric, each 2D profile contains the full 3D deformation profile.

The elastic theories of contact mechanics make predictions not only for the expected forces, but also for the substrate deformation profile as a function of sphere radius, contact radius, and elastic moduli \cite{Maugis1995}.
We plot these predictions with no free parameters as blue lines on the left side of Figure \ref{geo}(a).
For all deformations, we find that the elastic theory works well in the far field, but fails to describe the shape of the surface close to the contact line.
Even fitting with the Maugis theory by allowing the contact radius or sphere position to vary only does a marginally better job in describing the substrate deformation.

Even though the silicone meniscus below the sphere is solid, it bears a remarkable resemblance to a liquid capillary bridge.
Inspired by this similarity, we test how well a purely capillary theory describes this shape by fitting it with a surface of constant total curvature, $\kappa$, starting from the contact line \cite{jensen2015wetting}. 
We plot these fits as red lines on the right side of Figure \ref{geo}(a), extending the curves beyond the fit region to make clear where they begin to deviate from the data.

The pure capillary solution fits the measured surface profile extremely well close to the contact line, precisely where the elastic solution fails, but deviates from the measured profile in the far field, where the elastic solution works well.
We quantify the size of this domain of constant curvature, $\Delta S$, by measuring the path length along the profile from the contact line to the end of where the capillary solution fits well.
We plot $\Delta S$ and $-\kappa$ versus sphere displacement, $D$, over the entire stable contact regime in Figure \ref{geo}(b-c) for the same experiment as in Figure \ref{geo}(a) (black circles), as well as for an 18.8-$\mu$m-radius sphere (blue triangles) and a 32.0-$\mu$m-radius sphere (red squares).
The domain of constant curvature expands roughly exponentially with displacement, while simultaneously the magnitude of the curvature drops roughly exponentially.

We plot the sphere size dependence of the domain size, $\Delta S$, and the magnitude of the curvature, $-\kappa$,  at $D=0$ in Figure \ref{geo}(d-e).
Over a factor of three in particle radius, the initial inverse curvature (red squares) remains unchanged, while the domain of constant curvature (blue triangles) increases slightly.
The initial size of the capillary-dominated domain is $\Delta S_0 = 5.5 \pm 0.6$ $\mu$m, while the initial inverse curvature at $D = 0$ is $-1/\kappa_0 = 3.5 \pm 0.4$ $\mu$m.
These values are both comparable to the expected zero-force elastocapillary length: $\Upsilon_0/E= 3.6$ $\mu$m. 

We plot fitted values of length scales associated with the exponential growth/decay of the domain size (blue triangles) and curvature magnitude (red squares) on a log-log scale in Figure \ref{geo}(e).   
Both of these values display a roughly square-root dependence on the sphere radius over this range of sphere sizes.
Consequently, an additional length scale emerges from the dependence of the contact geometry on the sphere displacement.
By fitting the exponential length scales to a function of the form $L = \sqrt{lR}$, we obtain values for this new length scale of $l_{\Delta S} \approx 8.2$ $\mu$m and $l_\kappa \approx 5.2$ $\mu$m, both 1.5x larger than their corresponding initial length scales.

The contact profiles demonstrate a crossover from a capillary-dominated near field to an elastically-dominated far field, typical of elastocapillary behavior in soft materials. 
The transition between these domains is determined by the elastocapillary length,  which is usually assumed to be a material constant \cite{style2017elastocapillarity}.
Here, the dramatic increase of $\Delta S$ with sphere displacement suggests a concomitant increase in the elastocapillary length with deformation.
There are two ways that the elastocapillary length can grow with strain: either the elastic modulus drops, or the surface stress increases.
Bulk tensile tests rule out the former, showing instead moderate strain-stiffening at large strains.
Therefore, a growing elastocapillary length can only arise from a strain-dependent surface stress that increases with substrate surface deformation.

Inspired by these observations, we recently completed a complementary study directly measuring the strain-dependent surface stress of similar silicone gels, and found that it is indeed very sensitive to the surface strain of the material \cite{xujensen2017direct}.
In that case, we found that the strain dependence is described by a surface modulus, $\Lambda$, such that $\Upsilon(\epsilon) = \Upsilon_0 + \epsilon \Lambda$, where $\Lambda \approx 6 \Upsilon_0$. 
The surface modulus also introduces a new length scale, $\Lambda/E$, which for the silicone gels used in that study is about six times the zero-strain elastocapillary length.

\begin{figure}[t]
\includegraphics[width = 0.48\textwidth]{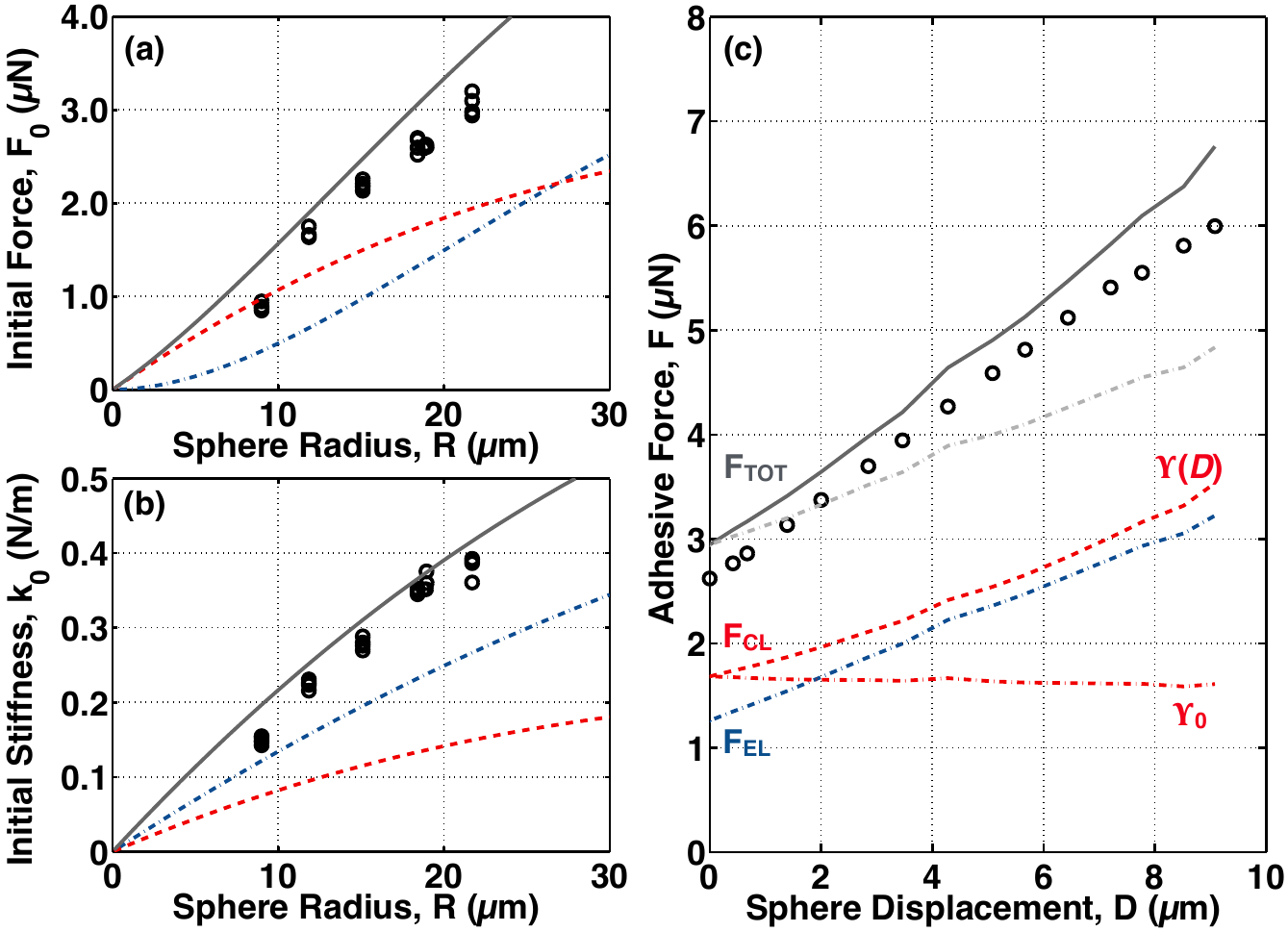}
\caption{\label{predict}Predicting adhesive forces.
\textit{(a-b)} Initial force, $F_0$, and initial contact stiffness, $k_0$, respectively, versus sphere radius, $R$. 
Plotted are: measured data (black symbols); predictions of elastic theory (blue dot-dashed lines) \cite{Maugis1995}; and capillary force predictions accounting for a strain-dependent surface stress (red dashed lines). 
Solid gray lines show the sum of the capillary and elastic predictions.
\textit{(c)} Measured force \emph{vs.} sphere displacement for an example experiment using an 18.9-$\mu$m-radius sphere (black points). 
Theoretical predictions are overlaid as: elastic theory (blue dot-dashed line); capillary prediction with constant or strain-dependent surface stress (red dot-dashed line and red dashed line, respectively); total estimated force with constant or strain-dependent surface stress (gray dot-dashed line and solid gray  line, respectively).
}
\end{figure}

Armed with these insights into the strain dependence of the surface stress, we revisit our estimate of capillary contributions to adhesive forces.
The complex strain state of these adhesion experiments makes a direct measurement of the surface modulus very difficult.
We therefore estimate the scaling of $\Upsilon$ with $D$ by taking the size of the domain of constant curvature as an approximate measure of the elastocapillary length.
Thus, we estimate the effective surface stress for a given deformation to be simply $\Upsilon \approx \Upsilon_0 \Delta S / \Delta S_0$.
Inserting this into Eq. \ref{cap} and using the mean values of the exponential fit parameters, we recalculate the capillary contributions to the total adhesive force over this range of sphere sizes, plotted as red dashed lines in Figure \ref{predict}(a-b).
As implemented, the strain dependence of the surface stress has no impact on the force at initial contact (Figure \ref{predict}(a)).
However, it significantly impacts the stiffness of the contact (Figure \ref{predict}(b)).
The sum totals of the elastic and strain-dependent capillary contributions are plotted as solid gray lines in Figure \ref{predict}(a-b).
Although these are simple calculations, they capture the magnitude and scaling of both adhesive force and contact stiffness over this range of particle sizes. 

This approach does remarkably well even at large deformations.
We plot the the measured force-displacement data for a single example experiment as black circles in Figure \ref{predict}(c).
For comparison, we calculate all of the variants of the force predictions, using the contact radius and growth of the constant curvature domain as measured from the images for this experiment:
the elastic prediction $F_{EL}$ \cite{Maugis1995} (blue dot-dashed line);
the capillary predictions $F_{CL}$ with both a fixed value of surface stress $\Upsilon = \Upsilon_0$ (red dot-dashed line) and with a strain-dependent surface stress $\Upsilon \approx \Upsilon_0 \exp{(D/L_{\Delta S})}$ (red dashed line);
the sum total forces, both using the fixed value of surface stress (gray dot-dashed line) and the strain-dependent surface stress (gray solid line).
The estimate with a fixed $\Upsilon = \Upsilon_0$ increasingly fails to describe the data as $D$ becomes large. 
However, the total force combining elastic and strain-dependent capillary contributions is in remarkable agreement with the measurements again despite the simplicity of our approach.

We have seen that theories of contact mechanics accounting only for bulk elasticity capture neither adhesive forces nor contact geometry in soft adhesion.
Rather, capillary forces arising from the surface stress of the compliant solid can contribute significantly to the total force.
However, simply including a fixed surface tension is not enough.  
Strain-dependent surface stresses are required to account for the structure and stiffness of soft adhesive contacts.
While our simple estimate of contact forces does a surprisingly good job, a complete elastocapillary theory of adhesion including strain-dependent surface stress needs to be developed. 
In particular, contributions from the interfacial curvature through the generalized Laplace-Young relation may need to be considered \cite{style2017elastocapillarity}.

\begin{acknowledgments}
We thank Dominic Vella, Desiree Plata, Anand Jagota, and Herbert Hui for useful discussions. 
We acknowledge funding from the National Science Foundation (CBET-1236086). 
\end{acknowledgments}


%

\end{document}